\documentclass{agujournal}
\usepackage[latin9]{inputenc}
\usepackage{array}
\usepackage{float}
\usepackage{booktabs}
\usepackage{url}
\usepackage{amsmath}
\usepackage{graphicx}

\makeatletter

\providecommand{\tabularnewline}{\\}

\@ifundefined{date}{}{\date{}}
\usepackage{amsfonts}
\usepackage{lineno}
\makeatother

\begin{document}

\title{Applications of physics-informed scientific machine learning in subsurface
science: A survey}

\authors{Alexander Y. Sun \affil{1}, Hongkyu Yoon \affil{2}, Chung-Yan
Shih \affil{3}, Zhi Zhong \affil{4}}

\affiliation{1}{Bureau of Economic Geology, Jackson School of
Geosciences, The University of Texas at Austin, Austin, TX} \affiliation{2}{Geomechanics
Department, Sandia National Laboratories, Albuquerque, NM} \affiliation{3}{Leidos,
Pittsburgh, PA} \affiliation{4}{School of Earth Resources, China
University of Geosciences, Wuhan, China}

\correspondingauthor{A. Y. Sun}{alex.sun@beg.utexas.edu} 

Geosystems are geological formations altered by humans activities
such as fossil energy exploration, waste disposal, geologic carbon
sequestration, and renewable energy generation. Geosystems also represent
a critical link in the global water-energy nexus, providing both the
source and buffering mechanisms for enabling societal adaptation to
climate variability and change. The responsible use and exploration
of geosystems are thus critical to the geosystem governance, which
in turn depends on the efficient monitoring, risk assessment, and
decision support tools for practical implementation. Large-scale,
physics-based models have long been developed and used for geosystem
management by incorporating geological domain knowledge such as stratigraphy,
governing equations of flow and mass transport in porous media, geological
and initial/boundary constraints, and field observations. Spatial
heterogeneities and the multiscale nature of geological formations,
however, pose significant challenges to the conventional numerical
models, especially when used in a simulation-based optimization framework
for decision support. Fast advances in machine learning (ML) algorithms
and novel sensing technologies in recent years have presented new
opportunities for the subsurface research community to improve the
efficacy and transparency of geosystem governance. Although recent
studies have shown the great promise of scientific ML (SciML) models,
questions remain on how to best leverage ML in the management of geosystems,
which are typified by multiscality, high-dimensionality, and data
resolution inhomogeneity. This survey will provide a systematic review
of the recent development and applications of domain-aware SciML in
geosystem researches, with an emphasis on how the accuracy, interpretability,
scalability, defensibility, and generalization skill of ML approaches
can be improved to better serve the geoscientific community.

\section{Introduction\label{sec:Introduction}}

Compartments of the subsurface domain (e.g., vadose zone, aquifer,
and oil and gas reservoirs) have provided essential services throughout
the human history. The increased exploration and utilization of the
subsurface in recent decades, exemplified by the shale gas revolution,
geological carbon sequestration, and enhanced geothermal energy recovery,
have put the concerns over geosystem integrity and sustainability
under unprecedented public scrutiny \citep{elsworth2016understanding,yeo2020causal}.
In this chapter, geosystems are defined broadly as the parts of lithosphere
that are modified directly or indirectly by human activities, including
mining, waste disposal, groundwater pumping and energy production
\citep{national2013induced}. Anthropogenic-induced changes may be
irreversible and have cascading social and environmental impacts (e.g.,
overpumping of aquifers may cause groundwater depletion, leading to
land subsidence and water quality deterioration, and increasing the
cost of food production). Thus, the sustainable management of geosystems
calls for integrated site characterization, risk assessment, and monitoring
data analytics that can lead to better understanding while promoting
inclusive and equitable policy making. Importantly, system operators
need to be able to explore and incorporate past experience and knowledge,
gained either from the same site or other similar sites, to quickly
identify optimal management actions, detect abnormal system signals,
and to prevent catastrophic events (e.g., induced seismicity and leakage)
from happening.

\begin{figure}
\includegraphics[width=5in]{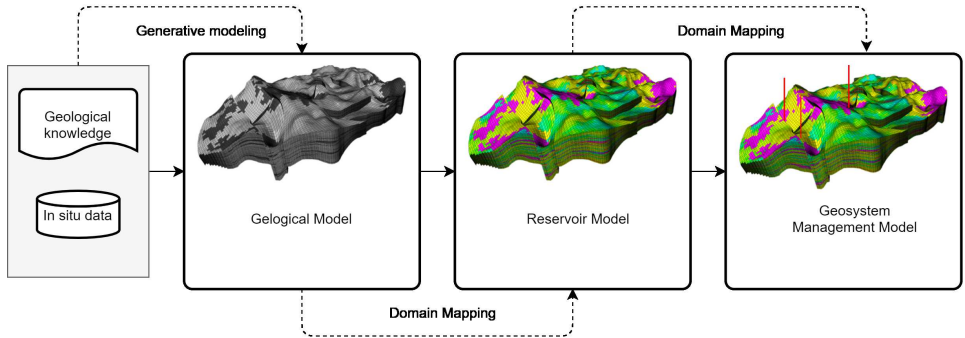}\caption{The conventional subsurface modeling workflow consists of (from left
to right) data collection and interpretation, geological modeling,
fluid modeling, and system modeling. Machine learning has potential
to automate all of these steps.\label{fig:1}}
\end{figure}

Fast advances in machine learning (ML) technologies have revolutionized
predictive and prescriptive analytics in recent years. Significant
interests exist in harnessing this new generation of ML tools for
Earth system studies \citep{reichstein2019deep,sun2019can,bergen2019machine}.
Unlike many other sectors, however, subsurface formations are often
poorly characterized and scarcely monitored, thus relying extensively
upon geological and geofluid modeling to generate spatially and temporally
continuous ``images'' of the subsurface. A conventional workflow
may consist of (a) geologic modeling, which seeks to provide a 3D
representation of the geosystem under study by fusing qualitative
interpretation of the geological structure, stratigraphy, and sedimentological
facies, as well as quantitative data on geologic properties; and (b)
fluid and geomechanical modeling, which describes the fluid flow,
mass transport, and formation deformations through physics-based governing
equations and the accompanying initial/boundary/forcing conditions.
Once established, the workflow is used to generate 3D ``images''
of the subsurface processes for inference and/or prediction (Figure
\ref{fig:1}).

We argue that subsurface modeling is inherently a semisupervised generative
modeling process \citep{chapelle2009semi}, in which joint data distributions
are learned via limited observations. A main difference is that in
field-scale subsurface modeling, observations are always sparse and
only indirectly related to data. For example, the observed quantities
may be well logs, but the data of interest are fluid saturation and
pore pressure (state variables); in this case, well logs are first
analyzed to infer stratigraphy and parameter distributions (e.g.,
porosity and permeability), which are then mapped to predictions of
the state variables. Physics-based modeling serves two purposes throughout
this process, namely, mapping the parameter space to state space,
and providing a spatiotemporal interpolation/extrapolation mechanism
that is guided by first principles and prior knowledge. A main issue
of this traditional workflow is that a significant amount of human
processing time and computational time is involved, limiting its efficiency
and potentially introducing significant latency and subjectivity during
the process. On the other hand, machines are good at automating processes
and learning proxy models after getting trained. Tremendous opportunities
now exist to integrate physics-based modeling and data-driven ML to
improve the accuracy and efficacy of the geoanalytics workflow.

To help the geoscientific community better embrace and incorporate
various ML methods, this chapter provides a survey of recent physics-based
ML developments in geosciences, with a focus on three main aspects---a
taxonomy of ML methods that have been used (Section \ref{sec:Categorization}),
brief introduction to some of the commonly used ML methods \ref{sec:algorithms}),
the types of use cases that are amenable to physics-based ML treatment
(Section \ref{sec:Application-Types} and Table \ref{tab:2}), and
finally challenges and future directions of ML applications in geosystem
modeling (Section \ref{sec:Challenges-and-Future}).

\section{Taxonomy of GeoML Methods \label{sec:Categorization}}

Table \ref{tab:1} lists the main notations and symbols used in this
survey.

\begin{table}
\noindent \begin{centering}
\caption{Notations and symbols used in this survey.\label{tab:1}}
\par\end{centering}
\begin{tabular}{p{3cm}p{6cm}}
\toprule 
Notations  & Descriptions \tabularnewline
\midrule 
${\bf {A}}$  & adjacency matrix \tabularnewline
${\bf {b}}$  & bias vector \tabularnewline
$d$  & number of features \tabularnewline
$\mathcal{D}$  & discriminative network \tabularnewline
${\bf {D}}$  & node degree matrix \tabularnewline
$\mathbb{E}$  & expectation operator \tabularnewline
$\mathcal{E}$  & graph edge set \tabularnewline
$\mathcal{G}$  & generative network \tabularnewline
${\bf {I}}$  & identity matrix \tabularnewline
${\bf {L}}$  & graph Laplacian \tabularnewline
$\mathcal{I}$  & prior information \tabularnewline
$n$  & number of samples \tabularnewline
$N$  & number of nodes in a graph\tabularnewline
$p(\cdot)$  & probability distribution \tabularnewline
$\mathcal{V}$  & graph node set \tabularnewline
${\bf {W}}$  & weight matrix \tabularnewline
${\bf {x}}\in{\mathbb{R}}^{d}$  & feature vector \tabularnewline
$\hat{{\bf {x}}}\in\mathbb{R}^{d}$  & estimated feature vector \tabularnewline
$X\in\mathbb{R}^{d}$  & feature variable \tabularnewline
$y$  & label, target variable, state variable \tabularnewline
${\bf {d}}^{obs}$  & data vector \tabularnewline
${\bf {z}}$  & latent space vector\tabularnewline
$\theta$  & parameters \tabularnewline
\bottomrule
\end{tabular}
\end{table}
Physics-based ML in the geoscientific domain (hereafter GeoML) is
a type of scientific machine learning (SciML) which, in turn, may
be considered a special branch of AI/ML that develops applied ML algorithms
for scientific discovery, with special emphases on domain knowledge
integration, data interpretation, cross-domain learning, and process
automation \citep{baker2019workshop}. A main thrust behind the current
SciML effort is to combine the strengths of physics-based models with
data-driven ML methods for better transparency, interpretability,
and explainability \citep{roscher2020explainable}. Unless otherwise
specified, we shall use the terms physics-based, process-based, and
mechanistic models interchangeably in this survey.

We provide three taxonomies of GeoML methods based on their design
and use. First, existing GeoML methods may be classified according
to the widely used ML taxonomy into unsupervised, supervised, and
reinforcement learning methods. In Figure \ref{fig:2}, this taxonomy
is used to group the existing GeoML applications, an exposition of
which will be deferred to Section \ref{sec:Application-Types}.

Another commonly used taxonomy is generative models vs. discriminative
models. Generative models seek to learn the probability distributions/patterns
of individual classes in a dataset, while discriminative models try
to predict the boundary between different classes in a dataset \citep{goodfellow2016deep}.
Thus, in a supervised learning setting and for given training samples
of input variables $X$ and label $y$, $\{({\bf x}_{i},y_{i})\}$,
a generate model learns the joint distribution $p(X,y)$ so that new
realizations can be generated, while a discriminative model learns
to predict the conditional distribution $p(y\mid X)$ directly. Generative
models can be used to learn from both labeled and unlabeled data in
supervised, unsupervised, or supervised tasks, while discriminative
models cannot learn from unlabeled data, but tend to outperform their
generative counterparts in supervised tasks \citep{chapelle2009semi}.
In GeoML, generative models are particularly appealing because of
the strong need for understanding the causal relationships, and because
the same underlying Bayesian frameworks are also employed in many
physics-based frameworks. In the classic Bayesian inversion framework,
for example, the parameter inference problem may be cast as \citep{sun2015model},
\begin{equation}
p(\theta\mid{\bf d}^{obs},\mathcal{I})=\frac{p(\theta\mid\mathcal{I})p({\bf d}^{obs}\mid\theta,\mathcal{I})}{p({\bf d}^{obs}\mid\mathcal{I})},\label{eq:1}
\end{equation}
where the posterior distribution of model parameters $\theta$ are
inferred from the state observations ${\bf d}^{obs}$ and prior knowledge
$\mathcal{I}$. Many physics-based ML applications exploit the use
of ML models for estimating the same distributions, but by fusing
domain knowledge to form priors and constraints.

On the basis of how physical laws and domain knowledge are incorporated,
existing GeoML methods fall into pre-training, physics-informed training,
residual modeling, and hybrid learning methods.

In pre-training methods, which are widely used in ML-based surrogate
modeling, prior knowledge and process-based models are mainly used
to generate training samples for ML from limited real information.
The physics is implicitly embedded in the training samples. After
the samples are generated, an ML method is then used to learn the
unknown mappings between parameters and model states through solving
a regression problem. In physics-informed training, physics laws and
constraints are utilized explicitly to formulate the learning problem,
such that ML models can reach a fidelity on par to PDE-based solvers.
Residual modeling methods use ML as a fine-tuning, post-processing
step, under the assumption that the process-based models reasonably
capture the large-scale ``picture'' but have certain missing processes,
due to either conceptual errors or unresolved/unmodeled processes
(e.g., subgrid processes). ML models are then trained to learn the
mapping between model inputs and error residuals (e.g., between model
outputs and observations), which are used to correct the effect of
missing processes on model outputs \citep{sun2019combining,reichstein2019deep}.
A main caveat of the existing ML paradigm is that models are trained
offline using historical data, and then deployed in operations, in
the hope that the future environment stays more or less under the
same conditions. This is referred to as the closed-world assumption,
namely, classes of all new test instances have already been seen during
training \citep{chen2018lifelong}. In some situations, new classes
of data may appear or the environment itself may drift over time;
in other situations, it is desirable to adapt a model trained on one
task to other similar tasks without training separate models. Hybrid
learning methods focus on continual or lifelong learning, in which
ML models and process-based models co-evolve to reflect new information
available. The past knowledge is accumulated and then used to help
future learning \citep{chen2018lifelong,parisi2019continual}. Hybrid
learning methods thus have elements of multitask learning, transfer
learning, and reinforcement learning from the ML side, and data assimilation
from the process modeling side. Understandably hybrid learning models
are more difficult to formulate and train, but they represent important
steps toward the ``real'' AI, in which agents learn to act reasonably
well not in a single domain but in many domains \citep{bostrom2014ethics}.

\section{Commonly Used GeoML Algorithms \label{sec:algorithms}}

For completeness, we briefly review the common algorithms and application
frameworks behind the GeoML use cases to be covered in Section \ref{sec:Application-Types}
(see also Table \ref{tab:2}). Most of the categories mentioned herein
are not exclusive. Autoencoders, the generative adversarial networks,
and graph neural networks are high-level ML algorithmic categories
that include many variants, while spatial-temporal methods and physics-informed
methods are application frameworks that may be implemented using any
of the ML methods.

\subsection{Autoencoders}

A main premise of the modern AI/ML is in representation learning,
which seeks to extract the low-dimensional features or to disentangle
the underlying factors of variation from learning subjects that can
support generic and effective learning \citep{bengio2013representation}.
An important class of methods for representation learning is autoencoders,
which are unsupervised learning methods that encode unlabeled input
data into low-dimensional embeddings (latent space variables) and
then reconstruct the original data from the encoded information. For
input data ${\bf x}$, the encoder maps it to a latent space vector,
${\bf z}=f({\bf x};{\bf W}_{e})$, while the decoder reconstructs
the input data, $\hat{{\bf x}}=g({\bf z};{\bf W}_{d}),$where ${\bf W}_{e}$
and ${\bf W}_{d}$ are weight matrices of the encoder and decoder,
respectively. The standard autoencoder is trained by minimizing the
reconstruction error, which implies a good representation should keep
the information of the input data well. Once trained, the autoencoder
may serve as a generative model (prior) for generating new samples,
clustering, or for dimension reduction. Variants of autoencoders include
variational autoencoder (VAE) and restricted Boltzmann machine (RBM)
\citep{goodfellow2016deep,doersch2016tutorial}.

It is worth pointing out that the notion of representation learning
has long been investigated in the context of parameterization and
inversion of physics-based models, although the primary goal there
is to make the inversion process less ill-posed by reducing the degree
of unknowns. In geosciences, autoencoders are closely related to stochastic
geological modeling, which is the main subject of study in geostatistics
\citep{journel1978mining}. In stochastic geological modeling, the
real geological formation is considered one realization of a generative
stochastic process that can only be ``anchored'' through a limited
set of measurements. The classic principle component analysis (PCA)
may be used to encode statistically stationary random processes, while
other algorithms, such as the multipoint statistics simulators, have
been commonly used to simulate more complex depositional environments
\citep{caers2004multiple,mariethoz2010direct}. Autoencoders, when
implemented using the deep convolutional neural nets (CNNs) \citep{chan2017parametrization,yoon2019permeability},
provide a more flexible tool for parameterizing the complex geological
processes and for generating (synthetic) training samples for the
downstream tasks, such as surrogate modeling. From this sense, autoencoders
fall in the category of pre-training methods.

\subsection{Generative adversarial networks}

The generative adversarial networks (GANs), introduced originally
in \citep{goodfellow2014generative}, have spurred strong interests
in geosciences. The vanilla GAN \citep{goodfellow2014generative}
trains a generative model (or generator) and a discriminator model
(discriminator) in a game theoretic setting. The generator $\mathcal{\hat{{\bf x}}=G}({\bf z};{\bf W}_{g})$
learns the data distribution and generates fake samples, while the
discriminator $\mathcal{D}({\bf x};{\bf W}_{d})$ predicts the probability
of fake samples being from the true data distribution, where ${\bf W}_{g}$
and ${\bf W}_{d}$ are trainable weight matrices. A minimax optimization
problem is formulated, in which the generator is trained to minimize
the reconstruction loss to generate more genuine samples, while the
discriminator is trained to maximize its probability of distinguishing
true samples from the fake samples, 
\begin{equation}
\text{arg}\underset{\mathcal{G}}{\min}\underset{\mathcal{D}}{\max}\mathbb{E}_{{\bf x}\sim p_{data}({\bf x})}\log[\mathcal{D}({\bf x};{\bf W}_{d})]+\mathbb{E}_{{\bf z}\sim p({\bf z})}\log[1-\mathcal{D}(\mathcal{G}({\bf z};{\bf W}_{g});{\bf W}_{d})],
\end{equation}
where $p_{data}(\cdot)$ and $p_{{\bf z}}(\cdot)$ are data and latent
variable distributions. In practice, the generator and discriminator
are trained in alternating loops, the weights of one model is frozen
when the weights of the other are updated. It has been shown that
if the discriminator is trained to optimality before each generator
update, minimizing the loss function is equivalent to minimizing the
Jensen-Shannon divergence between data $p_{data}(\cdot)$ and generator
$p_{{\bf \hat{x}}}(\cdot)$ distributions \citep{goodfellow2016deep}.

Many variants of the vanilla GAN have been proposed, such as the deep
convolutional GAN (DCGAN)\citep{radford2015unsupervised}, superresolution
GAN (SRGAN) \citep{ledig2017photo}, Cycle-GAN \citep{zhu2017unpaired},
StarGAN \citep{choi2018stargan}, and missing data imputation GAN
(GAIN) \citep{yoon2018gain}. Recent surveys of GANs are provided
in \citep{creswell2018generative,pan2019recent}. So far, GANs have
demonstrated superb performance in generating photo-realistic images
and learning cross-domain mappings. Training of the GANs, however,
are known to be challenging due to (a) larger-size networks, especially
those involving a long chain of CNN blocks and multiple pairs of generators/discriminators,
(b) the nonconvex cost functions used in GAN formulations, (c) diminished
gradient issue, namely, the discriminator is trained so well early
on in training that the generator's gradient vanishes and learns
nothing, and (d) the ``mode collapse'' problem, namely, the generator
only returns samples from a small number of modes of a mutimodal distribution
\citep{goodfellow2016deep}. In the literature, different strategies
have been proposed to alleviate some of the aforementioned issues.
For example, to adopt and modify deep CNNs for improving training
stability, the DCGAN architecture \citep{radford2015unsupervised}
was proposed by including stride convolutions and ReLu/LeakyRelu activation
functions in the convolution layers. To ameliorate stability issues
with the GAN loss function, the Wasserstein distance was introduced
in the Wasserstein GAN (WGAN) \citep{arjovsky2017wasserstein,gulrajani2017improved}
to measure the distance between generated and real data samples, which
was then used as the training criterion in a critic model. To remedy
the mode collapse problem, the multimodal GAN \citep{huang2018multimodal}
was introduced, in which the latent space is assumed to consist of
domain-invariant (called content code) and domain specific (called
style code) parts; the former is shared by all domains, while the
latter is only specific to one domain. The multimodal GAN is trained
by minimizing the image space reconstruction loss, and the latent
space reconstruction loss. In the context of continual learning, the
memory replay GAN \citep{wu2018memory} was proposed to learn from
a sequence of disjoint tasks. Like the autoencoders, GAN represents
a general formulation for supervised and semi-supervised learning,
thus its implementation is not restricted to certain types of network
models.

\subsection{Graph neural networks}

ML methods originating from the computer vision typically assume the
data has a Euclidean structure (i.e., grid like) or can be reasonably
made so through resampling. In many geoscience applications, data
naturally exhibits a non-Euclidean structure, such as the data related
to natural fracture networks and environmental sensor networks, or
the point cloud data obtained by lidar. These unstructured data types
are naturally represented using graphs. A graph $\mathcal{G}$ consists
of a set of nodes $\mathcal{V}$ and edges $\mathcal{E}$, $\mathcal{G}=(\mathcal{V},\mathcal{E})$.
Each node $v_{i}\in\mathcal{V}$ is characterized by its attributes
and has a varying number of neighbors, while each edge $e_{ij}\in\mathcal{E}$
denotes a link from node $v_{j}$ to $v_{i}$. The binary adjacency
matrix ${\bf A}$ is used to define graph connections, with its elements
$a_{ij}=1$ if there is edge between $i$ and $j$ and $a_{ij}=0$
otherwise.

Various graph neural networks (GNNs) have been introduced in recent
years to perform ML tasks on graphs, a problem known as ``geometric
learning'' \citep{bronstein2017geometric}. The success (e.g., efficiency
over deep learning problems) of CNN is owed to several nice properties
in its design, such as shift-invariance and local connectivity, which
lead to shared parameters and scalable networks \citep{goodfellow2016deep}.
A significant endeavor in the GNN development has been related to
extending these CNN properties to graphs using various clever tricks.

The graph convolutional neural networks (GCNN) extend CNN operations
to non-Euclidean domains and consist of two main classes of methods,
the spectral-based methods and the spatial-based methods. In spectral-based
methods, the convolution operation is defined in the spectral domain
through the normalized graph Laplacian, ${\bf L}$, defined as 
\begin{equation}
{\bf L}={\bf I}-{\bf D}^{-1/2}{\bf A}{\bf D}^{-1/2}={\bf U}{\bf \varLambda}{\bf U^{T}},
\end{equation}
where ${\bf A}$ is adjacency matrix, ${\bf I}$ is identify matrix,
${\bf D}$ is the node degree matrix (i.e, $d_{ii}=\sum_{j}a_{ij}$),
and ${\bf U}$ and $\Lambda$ are eigenvector matrix and diagonal
eigenvalue matrix of the normalized Laplacian. Utilizing ${\bf U}$
and $\Lambda$, the spectral graph convolution on input ${\bf x}$
is defined by a graph filter ${\bf g}_{\theta}$ \citep{bruna2013spectral}

\begin{equation}
{\bf g_{\theta}}*{\bf x}={\bf U}{\bf g}_{\theta}(\Lambda){\bf U}^{T}{\bf x},\label{eq:8}
\end{equation}
where $*$ denotes the graph convolution operator and the graph filter
${\bf g}_{\theta}(\Lambda)$ is parameterized by the learnable parameters
$\theta_{ij}$. Main limitations of the original graph filter given
in Eqn. \ref{eq:8} are it is non-local, only applicable to a single
domain (i.e., fixed graph topology), and involves the computationally
expensive eigendecomposition ($O(N^{3})$ time complexity) \citep{bronstein2017geometric,wu2020comprehensive}.
Later works proposed to make the graph filter less computationally
demanding by approximating ${\bf g}_{\theta}(\Lambda)$ using the
Chebychev polynomials of $\Lambda$, which led to ChebNet \citep{defferrard2016convolutional}
and Graph Convolutional Net (GCN) \citep{kipf2016semi}. It can be
shown these newer constructs lead to spatially localized filters,
such that the number of learnable parameters per layer does not depend
upon the size of the input \citep{bronstein2017geometric}. In the
case of GCN, for example, the following graph convolution operator
was proposed \citep{kipf2016semi}, 
\begin{equation}
{\bf g_{\theta}*x}\approx\theta\left({\bf I}+{\bf D}^{-1/2}{\bf A}{\bf D}^{-1/2}\right){\bf x}=\theta\left(\tilde{{\bf D}}^{-1/2}{\bf \tilde{A}}{\bf \tilde{D}}^{-1/2}\right){\bf x},\label{eq:5}
\end{equation}
where $\theta$ is a set of filter parameters, and a renormalization
trick was applied in the second equality in Eqn. \ref{eq:5} to improve
the numerical stability, ${\bf \tilde{A}=A+I}$ and $\tilde{d}_{ii}=\sum_{j}\tilde{a}_{ij}$.
The above graph convolutional operation can be generalized to multichannel
inputs ${\bf X}\in\mathbb{R}^{N\times C}$, such that the output is
given by $\tilde{{\bf D}}^{-1/2}{\bf \tilde{A}}{\bf \tilde{D}}^{-1/2}{\bf X}\Theta$,
where $\Theta$ is a matrix of filter parameters.

In spatial-based methods, graph convolution is defined directly over
a node's local neighborhood, instead via the eigendecomposition of
Laplacian. In diffusion CNN (DCNN), information propagation on a graph
is modeled as a diffusion process that goes from one node to its neighboring
node according to a transition probability \citep{atwood2016diffusion}.
The graph convolution in DCNN is defined as 
\begin{equation}
{\bf H}^{(k)}=\sigma\left({\bf W}^{(k)}\circ{\bf P}^{k}{\bf X}\right),\;k=1,\dots,K,
\end{equation}
where ${\bf X}$ is input matrix, ${\bf P}={\bf D}^{-1}{\bf A}$ is
transition probability matrix, $k$ defines the power of ${\bf P}$,
$K$ is the total number of power terms used (i.e., the number of
hops or diffusion steps) in the hidden state extraction, and ${\bf W}$
and ${\bf H}$ are the weight and hidden state matrices, respectively.
The final output is obtained by concatenating the hiddden state matrices
and then passing to an output layer. In GraphSAGE \citep{hamilton2017inductive}
and message passing neural network (MPNN) \citep{gilmer2017neural},
a set of aggregator functions are trained to learn to aggregate feature
information from a node's local neighborhood. In general, these networks
consist of three stages, message passing, node update, and readout.
That is, for each node $v$ and at the $k\text{-th}$ iteration, the
aggregation function $f_{k}$ combines the node's hidden representation
with those from its local neighbors $\mathcal{N}(v)$, which is then
passed to update functions to generate the hidden states for the next
iteration, 
\begin{equation}
{\bf h}_{v}^{(k)}=\sigma\left({\bf W}^{(k)},f_{k}\left({\bf h}_{v}^{(k-1)},\left\{ {\bf h}_{u}^{(k-1)},u\in\mathcal{N}(v)\right\} \right)\right),
\end{equation}
where ${\bf h}$ denotes a hidden-state vector. Finally, in the readout
stage, a fixed-length feature vector is computed by a readout function
and then passed to a fully connected layer to generate the outputs.

In general, spatial-based methods are more scalable and efficient
than the spectral methods because they do not involve the expensive
matrix factorization, the computation can be performed in mini-batches
and, more importantly, the local nature indicates that the weights
can be shared across nodes and structures \citep{wu2020comprehensive}.
Counterpart implementations of all well-established ML architectures
(e.g., GAN, autoencoder, and RNN) can now be found in GNNs. Recent
reviews on GNNs can be found in \citep{zhou2018graph,wu2020comprehensive}.

For subsurface applications, a main challenge is related to graph
formulation, namely, given a set of spatially discrete data, how to
connect the nodes. Common measures calculate certain pairwise distances
(e.g., correlation, Euclidean, city block), while other methods incorporate
the underlying physics (e.g., discrete fracture networks \citep{hyman2018identifying})
to identify the graphs.

\subsection{Spatiotemporal ML methods}

In this and the next subsection, we review two methodology categories
that use one or more of the aforementioned methods as construction
blocks. Spatiotemporal processes are omnipresent in geosystems and
represent an important area of study \citep{kyriakidis1999geostatistical}.
For gridded image-like data, the problem bears similarity to the video
processing problem in computer vision. In general, two classes of
ML methods have been applied, those involving only CNN blocks and
those combining with recurrent neural nets (RNNs).

Fully-connected CNNs can be used to model temporal dependencies by
stacking the most recent sequence of images/frames in a video stream.
In the simplest case, the channel dimension of the input tensor is
used to hold the sequence of images and CNN kernels are used to extract
features like in a typical CNN-based model (i.e., 2D kernels for a
stack of 2D images). In other methods, for example, C3D \citep{tran2015learning}
and temporal shift module (TSM) \citep{lin2019tsm}, an extra dimension
is added to the tensor variable to help extract temporal patterns.
C3D uses 3D CNN operators, which generally leads to much larger networks.
TSM was designed to shift part of the channels along the temporal
dimension, thus facilitating information extraction from neighboring
frames while adding almost no extra computational costs compared to
the 2D CNN methods \citep{lin2019tsm}.

The hybrid methods use a combination of RNNs with CNNs, using the
former to learn long-range temporal dependencies and the latter to
extract hierarchical features from each image. The convolutional long
short-term memory (ConvLSTM) network \citep{shi2015convolutional}
represents one the most well known methods under this category. In
ConvLSTM, features from convolution operations are embedded in the
LSTM cells, as described by the following series of operations \citep{shi2015convolutional}
\begin{eqnarray}
{\bf i}_{t} & = & \sigma\left({\bf W}_{xi}*{\bf X}_{t}+{\bf W}_{hi}*{\bf H}_{t-1}+{\bf W}_{ci}\circ{\bf C}_{t-1}+{\bf b}_{i}\right),\nonumber \\
{\bf f}_{t} & = & \sigma\left({\bf W}_{xf}*{\bf X}_{t}+{\bf W}_{hf}*{\bf H}_{t-1}+{\bf W}_{cf}\circ{\bf C}_{t-1}+{\bf b}_{f}\right),\nonumber \\
{\bf C}_{t} & = & {\bf f}_{t}\circ{\bf C}_{t-1}+{\bf i}_{t}\circ\tanh\left({\bf W}_{xc}*{\bf X}_{t}+{\bf W}_{hc}*{\bf H}_{t-1}+{\bf b}_{c}\right),\\
{\bf o}_{t} & = & \sigma\left({\bf W}_{xo}*{\bf X}_{t}+{\bf W}_{ho}*{\bf H}_{t-1}+{\bf W}_{co}*{\bf C}_{t}+{\bf b}_{o}\right),\nonumber \\
{\bf H}_{t} & = & {\bf o}_{t}\circ\tanh({\bf C}_{t}),\nonumber 
\end{eqnarray}
where ${\bf X}$, ${\bf H}$, and ${\bf C}$ are input, hidden, and
cell output matrices, ${\bf W}$ and ${\bf b}$ represent learnable
weights and biases, $\sigma$ and $\text{tanh}$ denote activation
functions, and ${\bf i}$, ${\bf f},$and ${\bf o}$ are the input,
forget, and output gates. The symbols $*$ and $\circ$ denote the
convolution operator and Hadamard (element-wise) product, respectively.
Because of its complexity and size, ConvLSTM networks may be more
difficult to train than the CNN-only methods.

Geological processes are known to exhibit certain correlation in space
and time. The convolution operations are like a local filter and not
good at catching large scale features, which is especially the case
for relatively shallow CNN-based models. In recent years, attention
mechanisms have been introduced to better capture the long-range dependencies
in space and time, and to give higher weight to most relevant information
\citep{vaswani2017attention}. In the location-based attention mechanism,
for example, input feature maps are transformed and used to calculate
a location-dependent attention map \citep{wang2018non,zhang2019self}
\begin{eqnarray}
{\bf F} & = & {\bf W}_{f}*{\bf X}_{i},\;{\bf G}={\bf W}_{g}*{\bf X}_{j},\\
\alpha_{ji} & = & \frac{\exp(s_{ij})}{\sum_{i=1}^{N}\exp(s_{ij})},\;s_{ij}={\bf F}^{T}{\bf G}\\
{\bf O}_{j} & = & \sum_{i=1}^{N}\alpha_{ji}({\bf W}_{h}*{\bf X}_{i}),
\end{eqnarray}
where ${\bf {\bf X}}\in\mathcal{\mathbb{R}}^{C\times H\times W}$
are the input feature maps, $C$, $H$, and $W$ are the channel and
spatial dimensions of the input feature map, $N=CW$ is the total
number of features in the feature map, ${\bf F}$ and ${\bf G}$ are
two transformed feature maps obtained by passing the inputs to separate
$1\times1$ convolutional layers, and the attention weights $\alpha_{ji}$
measure the influence of remote location $i$ on region $j$. The
resulting attention map is then concatenated with the input feature
maps to give the final outputs from the attention block. A similar
attention mechanism may be defined for the temporal dimension to catch
the temporal correlation \citep{zhu2018end}. The attention-based
ML models thus offer attractive alternatives to many parametric geostatistical
methods for 4D geoprocess modeling.

For unstructured data, GNNs can be used to learn spatial and temporal
relationships. For example, spatiotemporal graph convolution network
(ST-GCN) and spatiotemporal multi-graph convolution network \citep{geng2019spatiotemporal}
were used for skeleton-based action recognition and for ride share
forecast, respectively. The spatial-based GNNs may also be suitable
for the missing data problem, where the neighborhood information can
be used to estimate missing nodal values. For problems that can be
treated using GNNs, the resulting learnable parameter sizes are generally
much smaller.

\subsection{Physics-informed methods}

As mentioned in the last subsection, all GeoML applications that incorporate
certain domain knowledge or use process-based models in the workflow
may be considered physics informed. Recently, a number of SciML frameworks
have been developed to incorporate the governing equations in a more
principled way. In general, these methods may be divided into finite-dimensional
mapping methods, neural solver methods, and neural integral operator
methods \citep{li2020fourier}. All these methods seek to either parameterize
the solution of a PDE, $u=\mathcal{M}(a)$, where $\mathcal{M}$ is
model operator, $u\in\mathcal{U}$ is the solution and $a\in\mathcal{A}$
are parameters, or to approximate the model operator itself.

Finite-dimensional methods learn mappings between finite-dimensional
Euclidean spaces (e.g., the discretized parameter space and solution
space), which is similar to many use cases in computer vision. The
main difference is that additional PDE loss terms related to the PDE
being solved are incorporated. For example, \citet{zhu2019physics}
considered the steady-state flow problem in porous media (an elliptic
PDE) and used the variational form of the PDE residual as a loss term,
in addition to the data mismatch term. Many of the existing methods
(e.g., U-Net) from the computer vision can be directly applied in
these methods. A main limitation of the finite-dimensional methods
is they are grid specific (without resampling) and problem specific.

In neural solver methods, such as physics-informed neural networks
(PINNs) \citep{raissi2019physics,zhu2019physics,lu2019deepxde}, universal
differential equation (UDE) \citep{rackauckas2020universal}, and
PDE-Net \citep{long2019pde}, the neural networks (differentiable
functions by design) are used to approximate the solution and the
PDE residual is derived for the given PDE by leveraging auto differentiation
and neural symbolic computing. In general, these approaches assume
the PDE forms/classes are known a priori, although some approaches
(e.g., PDE-Net) can help to identify whether certain terms are present
in a PDE or not under relatively simple settings.

The neural integral operator methods \citep{fan2019multiscale,winovich2019convpde,li2020fourier}
parameterize the differential operators (e.g., Green's function, Fourier
transform) resulting from the solution of certain types of PDEs. These
methods are mesh independent and learn mappings between infinite-dimensional
spaces. In other words, a trained model has the ``super-resolution''
capability to map from a low-dimensional grid to a high-dimensional
grid.

All the physics-informed methods may provide accurate proxy models.
The advantages of these differential equation oriented methods are
(a) smooth solutions, by enforcing derivatives as constraints they
effectively impose smoothness in the solution, (b) extrapolation,
by forcing the NN to replicate the underlying differential equations
these methods also inherit the extrapolation capability of physics-based
models, which is lacking in purely data-driven methods, (c) closure
approximations, by parameterizing the closure terms using hidden neurons
they allow the unresolved processes to be represented and ``discovered''
in the solution process, and (d) less data requirements, which comes
as the result of the extensive constraints used in those frameworks.
On the other hand, the starting point of many methods are differential
equations, which means extensive knowledge and analysis are still
required to select and formulate the equations, a process that is
well known for its equifinality issue \citep{beven2001equifinality}.
Future works are still required to make the physics-informed methods
less PDE-class specific and be able to handle flexible initial/boundary
conditions and forcing terms.

\section{Applications \label{sec:Application-Types}}

\begin{figure}
\noindent \centering{}\includegraphics[width=5in]{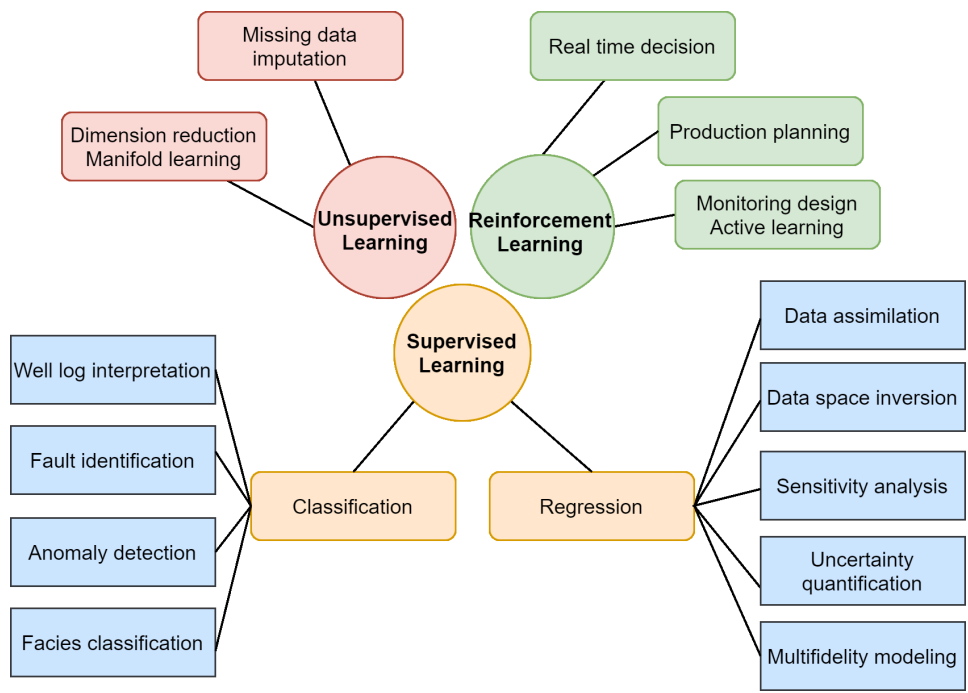}\caption{Common ML applications in geosciences may be grouped under unsupervised,
supervised, and reinforcement learning.\label{fig:2}}
\end{figure}

The number of GeoML publications has grown exponentially in recent
years. Here we review a selected set of recent GeoML applications
according to the taxonomy discussed under Section \ref{sec:Categorization},
and plotted in Fig. \ref{fig:2}. The list of publications is also
summarized in Table \ref{tab:2}, according to their ML model class,
model type, use case, and the way physics was incorporated. In making
the list, we mainly focused on reservoir-scale studies. Reviews of
porous flow ML applications in other disciplines (e.g., material science
and chemistry) can be found in \citep{alber2019integrating,brunton2020machine}.

Early works adopting the deep learning methods explored their strong
generative modeling capability for geologic simulation. In \citep{chan2017parametrization},
WGAN was used to generate binary facies realizations (bimodal). Training
samples were generated using a training image, which has long been
used in multipoint geostatistics as a geology-informed guide for constraining
image styles \citep{strebelle2002conditional}. WGAN was trained to
learn the latent space encoding of the bimodal facies field. The authors
showed that WGAN achieved much better performance than PCA, which
is a linear feature extractor that works best on single-modal, Gaussian-like
distributions. In \citep{liu2019deep}, a convolutional encoder was
trained to reconstruct a complex geologic model from its PCA parameterization.
A key idea there was to learn the mismatch between the naïve PCA representations
and the original high-fidelity counterparts such that new high-resolution
realizations can be generated using latent variables obtained from
PCA. Recently, DCGANs have been applied to generate drainage networks
by transforming the training network images to directional information
of flow on each node of the network \citep{kim2020connectivity}.
The generated network has been dramatically improved by optimal decomposition
of the drainage connectivity information into multiple binary layers
with the connectivity constraints stored.

A large number of GeoML applications fall under surrogate modeling,
which is not surprising given that the geocommunity has long been
utilizing surrogate models in model-based optimization, sensitivity
analysis, and uncertainty quantification \citep{forrester2009recent,razavi2012review}.
Because reservoir models are 2D or 3D distributed models, many ML
studies entailed some type of end-to-end, cross-domain learning architecture,
which translates an image of input parameter to state variable maps.
In general, these methods utilize physics-based porous flow models
to generate training samples. In \citep{mo2019deepco2}, a convolutional
autoencoder was trained to learn the cross-domain mapping between
permeability maps and reservoir states (pressure and saturation distributions)
at different times. In \citep{zhong2019predicting,zhong2020predicting},
a U-Net based convolutional GAN was trained to solve a similar problem.
Both studies also demonstrated the strong skill of ML-based models
in uncertainty quantification. In \citet{tang2020deep}, a hybrid
U-Net and ConvLSTM model was trained to learn the dynamic mappings
in multiphase reservoir simulations. In \citep{mo2020integration},
multi-level residual learning blocks were used to implement a GAN
model for surrogate modeling. ML techniques have also been combined
with model reduction techniques (e.g., proper orthogonal decomposition
or POD) to first reduce the dimension of models states before applying
ML \citep{jin2020deep}. The Darcy's flow problem has also been used
as a classic test case in many physics-informed studies, but generally
under relatively simple settings \citep{zhu2019physics,winovich2019convpde,li2020fourier}

Model calibration and parameter estimation represent an integral component
of the closed-loop geologic modeling workflow. A general strategy
has been using autoencoders to parameterize the model parameters as
random fields, the resulting latent variables are then ``calibrated''
using observation data in an outer-loop optimization, such as Markov
chain Monte Carlo \citep{laloy2018training} and ensemble smoothers
\citep{canchumuni2019towards,liu2019deep,mo2020integration,liu2020time}.
In Bayesian terms, this workflow yields the so-called conditional
realizations of the uncertain parameters, which are simply samples
of a posterior distribution informed by observations and priors (see
Eqn \ref{eq:1}). Other studies approached the inversion problem directly
using cross-domain mapping. For example, DiscoGAN and CycleGAN were
used to learn bidirectional \citep{sun2018discovering,wu2019inversionnet}
and tri-directional mappings \citep{zhonginversion}.

Many process-based models are high-dimensional and expensive to run,
prohibiting the direct use of cross-domain surrogate modeling. ML-based
multifidelity modeling offers an intermediate step. The general idea
is to reduce the requirements on high-fidelity model runs by utilizing
cheaper-to-run, lower fidelity models. Towards this goal, in \citep{perdikaris2017nonlinear},
a recursive Gaussian process was trained sequentially using data from
multiple model fidelity levels, reducing the number of high-fidelity
model runs. In \citep{meng2020composite}, a multifidelity PINN was
introduced to learn mappings (cross-correlations) between low- and
high-fidelity models, by assuming the mapping function can be decomposed
into a linear and a nonlinear part. Their method was expanded using
Bayesian neural networks to not only learn the correlation between
low- and high-fidelity data, but also give uncertainty quantification
\citep{meng2020multi}.

Fractures and faults are extensively studied in geosystem modeling
for risk assessment and production planning. In \citep{schwarzer2019learning},
a recurrent GNN was used to predict the fracture propagation, using
simulation samples from high-fidelity discrete fracture network models.
In \citep{Sidorov_2019_ICCV}, GNN was used to extract crack patterns
directly from high-resolution images, which may have a significant
implication to a wide range of geological applications.

Ultimately, the goal of geosystem modeling is to train ML agents to
quickly identify optimal solutions and/or policies, which is a challenging
problem that requires integrating many pieces in the current ML ecosystems.
The recently advanced deep reinforcement learning algorithms offer
a new paradigm for exploiting past experiences while exploring new
solutions \citep{mnih2013playing}. In general, model-based deep reinforcement
learning frameworks solve a sequential decision making problem. At
any time, the agent chooses a trajectory that maximizes the future
rewards. Doing so would require hopping many system states in the
system space, which is challenging for high-dimensional systems. In
\citep{sun2020optimal}, the deep Q learning (DQL) algorithm was used
to identify the optimal injection schedule in a geologic carbon sequestration
planning. A deep surrogate, autoregressive model was trained using
U-Net to facilitate state transition prediction. A discrete planning
horizon was assumed to reduce the total computational cost. In \citep{ma2019waterflooding},
a set of deep reinforcement learning algorithms were applied to maximize
the net present value of water flooding rates in oil reservoirs. It
remains a challenge to generate the surrogate models for arbitrary
state transitions, such as that is encountered in well placement problems
where both the number and locations of new wells need to be optimized.

\begin{table}[H]
\caption{List of physics-informed GeoML applications\label{tab:2}}

\begin{tabular}{p{2cm}p{2cm}p{3cm}p{3cm}p{2cm}}
\toprule 
Model class  & Model Name  & Use Case  & Physics Used  & Citation\tabularnewline
\midrule 
Autoencoder  & Conv-VAE  & Geologic simulation  & Geology  & \citep{laloy2018training}, \citep{canchumuni2019towards} \tabularnewline
 & CNN-PCA  & Geologic simulation  & Geology, reservoir model  & \citep{liu2019deep} \tabularnewline
 & Conv-AE  & Surrogate modeling  & Reservoir model outputs  & \citep{mo2019deepco2},\citep{mo2019deep} \tabularnewline
 & CNN  & Model reduction, surrogate modeling  & Multiphase flow model  & \citep{wang2020efficient} \tabularnewline
\midrule 
Generative adversarial networks  & WGAN  & Unconditional geologic simulation  & Geology  & \citep{chan2017parametrization} \tabularnewline
 & DiscoGAN  & Bidirectional parameter-state mapping  & Groundwater model outputs & \citep{sun2018discovering} \tabularnewline
 & Conditional GAN  & Surrogate modeling  & Reservoir model outputs  & \citep{zhong2019predicting},\citep{zhong2020predicting} \tabularnewline
 & ConvGAN  & Inversion  & Geology, groundwater model  & \citep{mo2020integration} \tabularnewline
 & CycleGAN  & Tridirectional parameter-state mapping  & Multiphase flow, petrophysics  & \citep{zhonginversion} \tabularnewline
 & CycleGAN  & Bidirectional parameter-state mapping  & Full waver inversion  & \citep{wu2019inversionnet},\citep{wang2019seismic}\tabularnewline
 & DCGAN  & Drainage networks  & Hydraulic connectivity  & \citep{kim2020connectivity} \tabularnewline
\midrule 
Graph neural nets  & Diffusion GNN  & Discrete fracture modeling  & Fracture connectivity & \citep{schwarzer2019learning}, \citep{Sidorov_2019_ICCV} \tabularnewline
\midrule 
Spatiotemporal  & Unet-LSTM  & Surrogate modeling  & Reservoir model outputs  & \citep{tang2020deep} \tabularnewline
 & CNN only  & Surrogate modeling  & Reservoir model  & \citep{zhong2019predicting}, \citep{mo2019deepco2} \tabularnewline
 &  &  &  & \tabularnewline
\midrule 
PDE-informed  & DNN  & Parameter estimation  & Soil physics  & \citep{tartakovsky2020physics} \tabularnewline
 & DNN  & Forward modeling and inversion  & Soil physics  & \citep{bekele2020physics} \tabularnewline
 & Conv-AE  & Surrogate modeling  & Flow equation  & \citep{zhu2019physics} \tabularnewline
 & DNN  & Immsicible flow modeling  & Reservoir flow equation  & \citep{tchelepi2020limitations} \tabularnewline
 & Neural operator  & Surrogate modeling  & Darcy flow equation  & \citep{li2020fourier} \tabularnewline
\bottomrule
\end{tabular}
\end{table}

\section{Challenges and Future Directions \label{sec:Challenges-and-Future}}

Our survey shows that GeoML has opened a new window for tackling longstanding
problems in geological modeling and geosystem management. Nevertheless,
a number of challenges remain, which are described below.

\subsection*{Training data availability}

GeoML tasks require datasets for training, validation, and testing.
In the subsurface domain, data acquisition can be costly. For example,
to acquire 3D seismic data, an operator may spend at least \$1M before
seeing results. The costs of drilling new exploration wells are on
the same order of magnitudes. Thus, data augmentation using synthetic
datasets will play an important role in improving the current generalization
capability of ML models. A main challenge is related to generating
realistic datasets that also meet unseen field conditions. In addition,
generating simulation data for subsurface applications can also be
time consuming if the parameter search space is large and requires
substantial computational resources.

Efforts from the public and private sectors have started to make data
available. Government agencies encourage or require oil and gas operators
to regularly report well information (e.g., drilling, completion,
plugging, production, etc.) and make the data available to the public.
Standing on the foundation, companies integrate the data and added
proprietary assessments for commercial licenses. The U.S. Energy Information
Administration (EIA) implements multiple approaches to facilitate
data access (\url{https://www.eia.gov/opendata}). Over the last decade,
the National Energy Technology Laboratory (NETL) has developed a data
repository and laboratory, called Energy Data eXchange (EDX), to curate
and preserve data for reuse and collaboration that supports the entire
life cycle of data (\url{https://edx.netl.doe.gov/}). Open Energy
Information (\url{https://openei.org}) represents an example of community-driven
platform for sharing energy data. However, challenges related to data
Findability, Accessibility, Interoperability, and Reuse (FAIR) remain
to be solved. For example, government agencies or data publishers
may have different definitions and data capturing processes. Comparing
data on the same basis requires additional processing and deciphering
\citep{shih2021}.

Because of the high cost of acquiring data, proprietary, and other
reasons, companies often are hesitant to share their data to form
a unified or centralized dataset for ML training. A new approach,
federated learning, is emerging to address the data privacy issue
\citep{konevcny2016federated}. The approach trains an algorithm across
decentralized models with their local datasets. Instead of sending
the data to form a unified dataset, federated learning exchanges parameters
among local models to generate a global model. This approach shows
one way to solve the data issues in the subsurface fields to promote
collaboration.

\subsection*{Model scalibility}

Geosystems are a type of high-dimensional dynamic systems. Image-based,
deep learning algorithms originating from computer vision were developed
for fixed, small-sized training images. Large-scale models (e.g.,
hyperresolution groundwater model) are thus too big to use without
resampling, a procedure that inevitably loses fine details. There
is a strong need for developing multi-resolution, multi-fidelity ML
models that are suitable for uncovering multiscale geological patterns.
We are beginning to see new developments in this direction from the
applied mathematics \citep{park2020multiresolution,meng2020multi,li2020fourier,fan2019multiscale}.
However, the feasibility of these approaches on field-scale problems
in geosciences needs to be tested.

From the cyber-infrastructure side, next-generation AI/ML acceleration
hardware continuously evolve to tackle the scalability issue. For
example, a recent pilot study in computational fluid dynamics showed
that it could be more than 200 times faster than the same workload
on an optimized number of cores on the NETL's supercomputer JOULE
2.0 \citep{rocki2020fast}. Similar scaling performance has been reported
on other exascale computing clusters involving hundreds of GPU's \citep{byna2020exahdf5}.

\subsection*{Domain transferrability }

Even though geologic properties are largely static, the boundary and
forcing conditions of geosystems are dynamic. A significant challenge
is related to adapting ML models trained for one set of conditions
or a single site (single domain) to other conditions (multiple domains),
with potentially different geometries and boundary/forcing conditions.
This problem has been tackled under lifelong learning (see Section
\ref{sec:Categorization}). In recent year, few-shot meta-learning
algorithms \citep{finn2017model,sun2019meta} have been developed
to enable domain transferrability. The goal of meta-learning is to
train a model on a variety of learning tasks such that the trained
model can discover the common structure among tasks (i.e., learning
to learn), which is then used solve new learning tasks using only
a small number of training samples \citep{finn2017model}. Future
GeoML research needs to adapt these new developments to enhance transfer
learning across geoscience domains.

\section{Conclusions}

Geosystems play an important role in the current societal adaptation
to climate change. Tremendous opportunities exist in applying AI/ML
to manage the geosystems in transparent, fair, and sustainable ways.
This chapter provided a review of the current applications and practices
of ML in the geosystem management. Significant progress has been made
in recent years to incorporate deep learning algorithms and physics-based
learning. Nevertheless, many of the current approaches/models are
limited by their generalization capability because of data limitations,
domain specificity, and/or resolution limitation. In addition, many
of the current models were demonstrated over simplistic toy problems.
Future efforts should focus on mitigating these aspects to make GeoML
models more generalizable and trustworthy.

\section*{Acknowledgments}

A. Sun was partly supported by the U.S. Department of Energy, National
Energy Technology Laboratory (NETL) under grants DE-FE0026515, DE-FE0031544,
and the Science-informed Machine Learning for Accelerating Real-Time
Decisions in Subsurface Applications (SMART) Initiative. This work
was also supported by the Laboratory Directed Research and Development
LDRD program at Sandia National Laboratories (213008). Sandia National
Laboratories is a multimission laboratory managed and operated by
National Technology and Engineering Solutions of Sandia, LLC., a wholly
owned subsidiary of Honeywell International, Inc., for the U.S. Department
of Energy's National Nuclear Security Administration under contract
DE-NA-0003525. This paper describes objective technical results and
analysis. Any subjective views or opinions that might be expressed
in the paper do not necessarily represent the views of the U.S. Department
of Energy or the United States Government.

\bibliographystyle{elsarticle-harv}
\bibliography{C:/Alex/deeplearn/AnujBookChapter/chapter1latex/gan_review}

\end{document}